\documentclass{moriondNOTITLES}

\usepackage{xspace}
\usepackage{amsmath}
\usepackage{units}

\def\be{\begin{equation}}
\def\ee{\end{equation}}
\def\bea{\begin{eqnarray}}
\def\eea{\end{eqnarray}}

\newcommand{\PH}{\ensuremath{\mathrm{H}}\xspace}
\newcommand{\fb}{\unit{fb}}
\newcommand{\TeV}{\unit{TeV}}
\newcommand{\GeV}{\unit{GeV}}

\newcommand{\PV}{\ensuremath{\mathrm{V}}\xspace}
\newcommand{\PW}{\ensuremath{\mathrm{W}}\xspace}
\newcommand{\PZ}{\ensuremath{\mathrm{Z}}\xspace}
\newcommand{\PQc}{\ensuremath{\mathrm{c}}\xspace}
\newcommand{\PAQc}{\ensuremath{\mathrm{\bar{c}}}\xspace}
\newcommand{\PQb}{\ensuremath{\mathrm{b}}\xspace}
\newcommand{\PAQb}{\ensuremath{\mathrm{\bar{b}}}\xspace}

\newcommand{\ttbar}{\ensuremath{\mathrm{t\bar{t}}}\xspace}
\newcommand{\bbbar}{\ensuremath{\mathrm{b\bar{b}}}\xspace}
\newcommand{\ccbar}{\ensuremath{\mathrm{c\bar{c}}}\xspace}
\newcommand{\fbinv}{\ensuremath{\fb^{-1}}\xspace}

\newcommand{\ttH}{\ensuremath{\ttbar\PH}\xspace}
\newcommand{\hcc}{\ensuremath{\PH{\to}\PQc\PAQc}\xspace}
\newcommand{\hbb}{\ensuremath{\PH{\to}\PQb\PAQb}\xspace}

\newcommand{\brHcc}[1][{}]{\ensuremath{\mathcal{B}_{\text{#1}}(\PH{\to}\PQc\PAQc)}\xspace}
\newcommand{\brHbb}[1][{}]{\ensuremath{\mathcal{B}_{\text{#1}}(\PH{\to}\PQb\PAQb)}\xspace}

\newcommand{\yc}{\ensuremath{y_c}\xspace}
\newcommand{\kappaC}{\ensuremath{\kappa_{\PQc}}\xspace}
\newcommand{\kappaB}{\ensuremath{\kappa_{\PQb}}\xspace}

\newcommand{\ttZ}{\ensuremath{\ttbar\PZ}\xspace}

\newcommand{\VH}{\ensuremath{\PV\PH}\xspace}

\newcommand{\zcc}{\ensuremath{\PZ{\to}\ccbar}\xspace}
\newcommand{\zbb}{\ensuremath{\PZ{\to}\bbbar}\xspace}
\newcommand{\ttHbb}{\ensuremath{\ttH(\hbb)}\xspace}
\newcommand{\ttHcc}{\ensuremath{\ttH(\hcc)}\xspace}
\newcommand{\ttZbb}{\ensuremath{\ttZ(\zbb)}\xspace}
\newcommand{\ttZcc}{\ensuremath{\ttZ(\zcc)}\xspace}

\newcommand{\sigstrength}[1]{\ensuremath{\mu_{#1}}\xspace}
\newcommand{\muHcc}{\ensuremath{\sigstrength{\ttHcc}}\xspace}
\newcommand{\muJustHcc}{\ensuremath{\sigstrength{\hcc}}\xspace}
\newcommand{\muHbb}{\ensuremath{\sigstrength{\ttHbb}}\xspace}
\newcommand{\muZcc}{\ensuremath{\sigstrength{\ttZcc}}\xspace}
\newcommand{\muZbb}{\ensuremath{\sigstrength{\ttZbb}}\xspace}

\newcommand{\jets}{\text{jets}}
\newcommand{\ttjets}{\ensuremath{\ttbar{+}\jets}\xspace}

\newcommand{\ttlight}{\ensuremath{\ttbar{+}\text{light}}\xspace}

\newcommand{\ttbb}{\ensuremath{\ttbar{+}{\geq}2\PQb}\xspace} % tt + >=2 b-jets
\newcommand{\ttbj}{\ensuremath{\ttbar{+}\PQb}\xspace} % tt + =1 b-jets
\newcommand{\ttcc}{\ensuremath{\ttbar{+}{\geq}2\PQc}\xspace} % tt + >=2 c-jets
\newcommand{\ttcj}{\ensuremath{\ttbar{+}\PQc}\xspace} % tt + =1 c-jets
\newcommand{\deepJet}{\textsc{DeepJet}\xspace}
\newcommand{\pNet}{{ParticleNet}\xspace}
\newcommand{\parT}{{ParT}\xspace}

\newcommand{\FH}{\ensuremath{\text{0L}}\xspace}
\newcommand{\SL}{\ensuremath{\text{1L}}\xspace}
\newcommand{\DL}{\ensuremath{\text{2L}}\xspace}

\newcommand{\PB}{\ensuremath{\mathrm{B}}\xspace}
\newcommand{\PQC}{\ensuremath{\mathrm{C}}\xspace}
\newcommand{\pBpC}{\ensuremath{p_{\PB{+}\PQC}}\xspace}
\newcommand{\pBvC}{\ensuremath{p_{\PB\textrm{vs}\PQC}}\xspace}

\newcommand{\pmasym}[2]{\ensuremath{\,^{#1}_{#2}}}

\newcommand{\Photo}{}

\begin{document}
\vspace*{4cm}
\title{Search for Higgs boson decay to a charm quark-antiquark pair via \ttH production}

\author{Sebastian Wuchterl\\on behalf of the CMS Collaboration\footnote{Copyright [2024] CERN for the benefit of the CMS Collaboration. Reproduction of this article or parts of it is allowed as specified in the CC-BY-4.0 license}}

\address{CERN, Espl. des Particules 1,\\
1211 Meyrin, Switzerland}

\maketitle\abstracts{
In these proceedings, a search for the standard model Higgs boson decaying to a charm quark-antiquark pair, \hcc, produced in association with a top quark-antiquark pair (\ttH) is presented.
The search is performed using proton-proton collision data collected by the CMS experiment at ${\sqrt{s}=13}\,\TeV$, corresponding to an integrated luminosity of $138\,\fbinv$.
The Higgs boson decay to a bottom quark-antiquark pair is measured simultaneously and the observed \ttHbb event rate relative to the standard model expectation is found to be $0.91\pmasym{+0.26}{-0.22}$.
The observed (expected) upper limit at $95\%$ confidence level (CL) for \ttHcc production is 7.8 ($8.7$) times the standard model prediction.
Combined with a previous search for \hcc via associated production with a \PW or \PZ boson, the observed (expected) $95\%$ CL interval on the Higgs-charm Yukawa coupling modifier, $\kappaC$, is $|\kappaC|<3.5$ ($|\kappaC|<2.7$).}

\section{Introduction}

The discovery of a Higgs boson (\PH) by the ATLAS~\cite{bib:ATLAS} and CMS~\cite{bib:CMS} experiments in 2012~\cite{Aad:2012tfa,CMS:HIG-12-028} marks a milestone in understanding the electroweak symmetry breaking.
With a measured mass of $125.38\pm0.14\,\GeV$~\cite{CMS:HIG-19-004}, the observed interactions with gauge bosons and third-generation fermions, as well as all measured properties of the Higgs boson, agree with standard model (SM) predictions~\cite{CMS:HIG-22-001,ATLAS:2022vkf}.
Following the first evidence of Higgs boson decays to muons~\cite{CMS:HIG-17-019}, an important next step is the observation of the Higgs boson coupling to second-generation quarks.
Searches for Higgs boson decays to a charm quark-antiquark pair (\ccbar) provide direct access to the charm Yukawa coupling (\yc).
Using $138\,\fbinv$ of proton-proton collision data at $13\,\TeV$, the ATLAS~\cite{ATLAS:2024yzu} and CMS~\cite{CMS:HIG-21-008} Collaborations reported observed (expected) 95\% confidence level (CL) intervals on $\kappaC=\yc/{\yc}^{\text{SM}}$ of $|\kappaC|<4.2$ (4.1) and ${1.1 <|\kappaC|<5.5}$ ($|\kappaC|<3.4$), respectively, studying the associated production of a Higgs boson with a \PV (\PW or \PZ) boson.
These proceedings report a new search for \hcc via associated production of a Higgs boson with a top quark-antiquark pair (\ttH), while the \hbb decay is measured simultaneously. In addition, the SM processes \ttZcc and \ttZbb are measured to validate the analysis strategy for \ttH production.

\section{Analysis strategy}
\begin{figure}
\centerline{\includegraphics[width=0.5\linewidth]{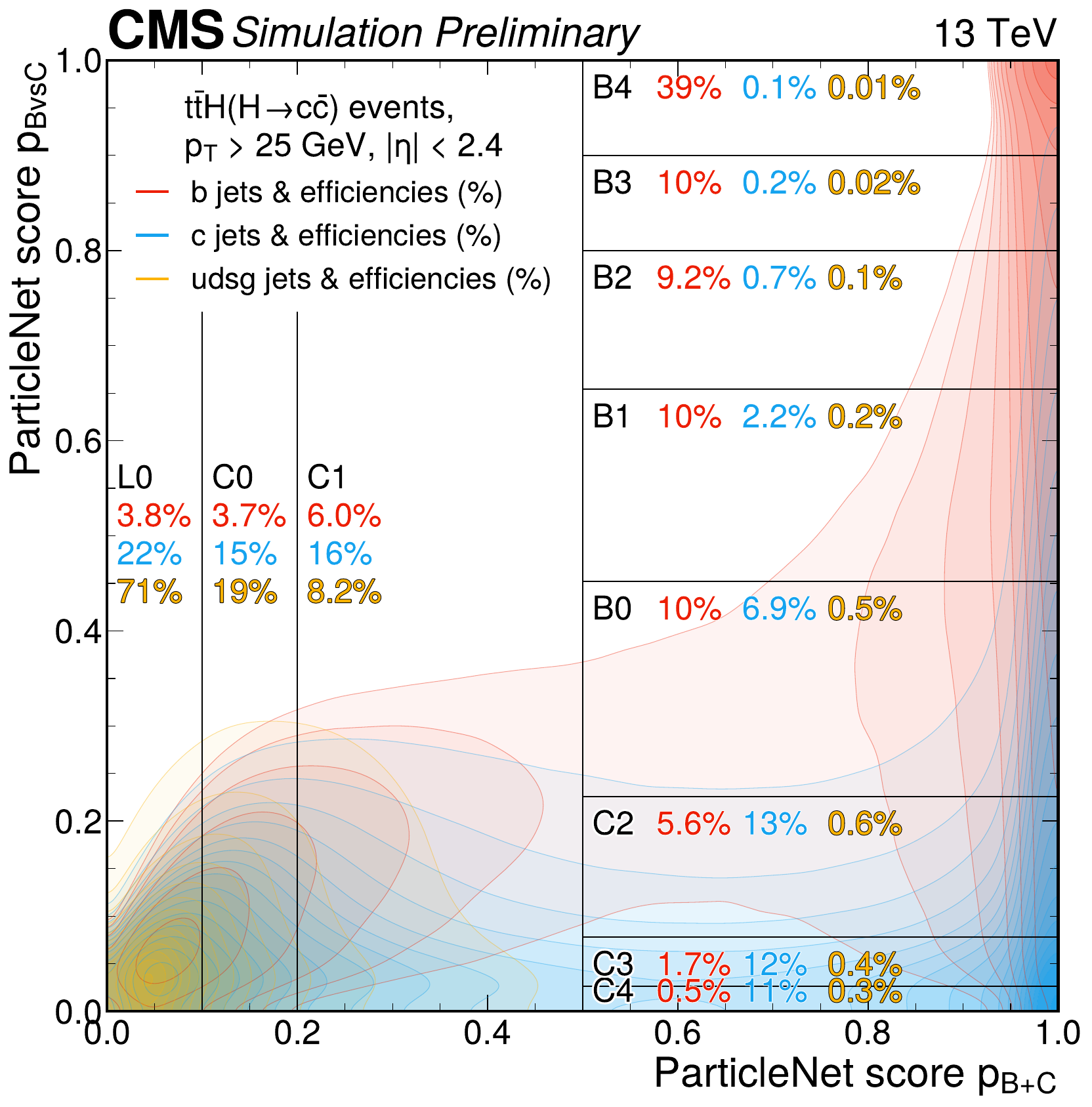}}
\hfill
\caption[]{Distribution of \PQb, \PQc and light-flavor jets in the two-dimensional \pNet discriminant plane, together with the tagging categories and the corresponding efficiencies~\cite{CMS-PAS-HIG-24-018}.}
\label{fig:ftag}
\end{figure}

The analysis uses proton-proton collision data at $\sqrt{s}=13\,\TeV$, collected by the CMS detector and corresponding to an integrated luminosity of $138\,\fbinv$ and is carried out in three channels targeting the fully hadronic (\FH), semileptonic (\SL), and dileptonic (\DL) decays of the \ttbar pair.
To identify the jet flavor, the \pNet algorithm~\cite{Qu:2019gqs} is used.
Two discriminants are defined from the \pNet outputs: \pBpC, which separates heavy- from light-flavor jets, and \pBvC, which distinguishes \PQb from \PQc jets.
Based on these, 11 mutually exclusive tagging categories are defined as shown in Fig.~\ref{fig:ftag}.
Compared to the previous tagging algorithm used in the CMS Collaboration, \deepJet~\cite{Bols:2020bkb}, \pNet improves the \PQb vs. light and \PQc vs. \PQb jet rejection by up to a factor of two each at the same signal jet efficiency.

A multiclass event classifier based on \parT~\cite{Qu:2022mxj} is developed to classify events. The \parT classifier is a transformer based algorithm which combines inputs such basic kinematic properties and tagging information with pairwise features for all object pairs derived using their four-momenta to better learn the event kinematic properties and object correlations.
The classifier is trained to assign likelihood scores across 10 (9) classes in the \FH (\SL, \DL) channel: two \ttH classes (\ttHcc, \ttHbb), two \ttZ classes (\ttZcc, \ttZbb), five \ttjets classes (\ttbj, \ttbb, \ttcj, \ttcc, \ttlight), and only in the \FH channel, one class for QCD multijet.

The output scores of the \parT classifier are used to optimize the event selection and categorization.
In the \FH channel, a stringent requirement on the QCD multijet score suppresses this background by about four orders of magnitude, making it negligible compared to \ttjets.
Similarly, a selection on the \ttlight discriminant further reduces \ttlight contamination in all channels.
Only events with a high \ttH or \ttZ score value are retained for the signal extraction and background estimation.
Events passing these criteria are categorized into four signal regions (SRs), enriched in \ttHcc, \ttHbb, \ttZcc, and \ttZbb, together with five control regions (CRs), to estimate the normalizations of the \ttjets background components.

The \ttH and \ttZ production rates are determined via a binned profile likelihood fit to data.
For each of the four signal processes, the expected yield is scaled by a signal strength modifier $\mu$, defined as ${\left(\sigma\mathcal{B}\right)_{\text{obs}}/\left(\sigma\mathcal{B}\right)_{\text{SM}}}$ where $\sigma$ is the production cross section and $\mathcal{B}$ is the branching fraction of the \PH or \PZ decays.
The dominant background, \ttjets, is estimated with the normalizations of the \ttcj, \ttcc, \ttbj, \ttbb, and \ttlight components floating independently and constrained in dedicated control regions.
For \muHcc, the leading uncertainty is statistical (${\approx}74\%$), whereas for \muHbb, the largest uncertainties are theoretical in the \ttjets model, representing ${\approx}60\%$ of the total uncertainty.

\section{Results}

The analysis is validated by measuring the \ttZ signal strengths $\muZcc = 1.02\pmasym{+0.79}{-0.84}$ and $\muZbb = 1.47\pmasym{+0.45}{-0.41}$,
which agree with the SM prediction within one (two) standard deviation for the \zcc (\zbb) decay mode, and with measurements in the leptonic decay channel~\cite{CMS:TOP-23-004,ATLAS:2023eld}.

\begin{figure}
\begin{minipage}{0.49\linewidth}
\centerline{\includegraphics[width=0.99\linewidth]{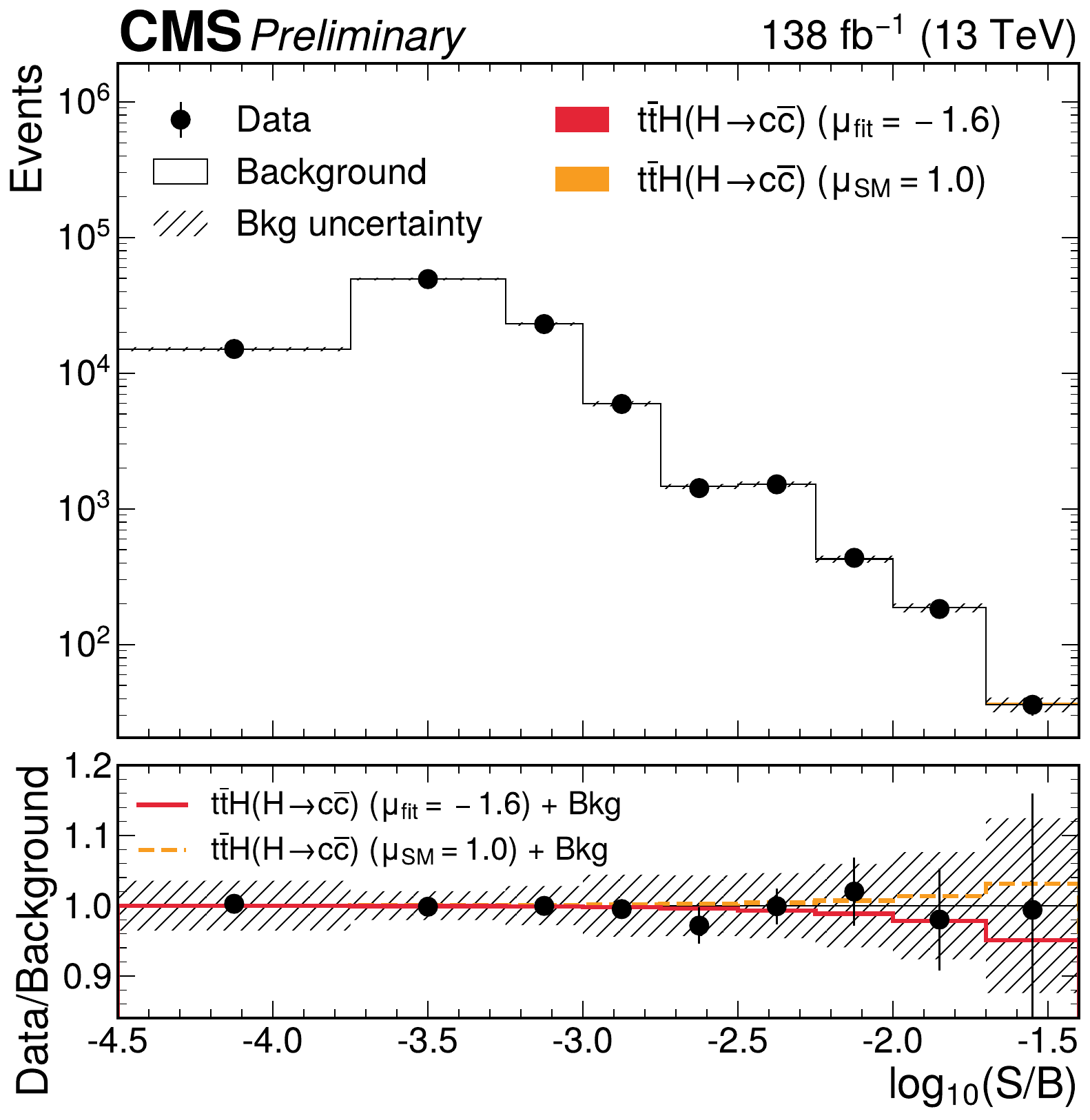}}
\end{minipage}
\hfill
\begin{minipage}{0.49\linewidth}
\centerline{\includegraphics[width=0.99\linewidth]{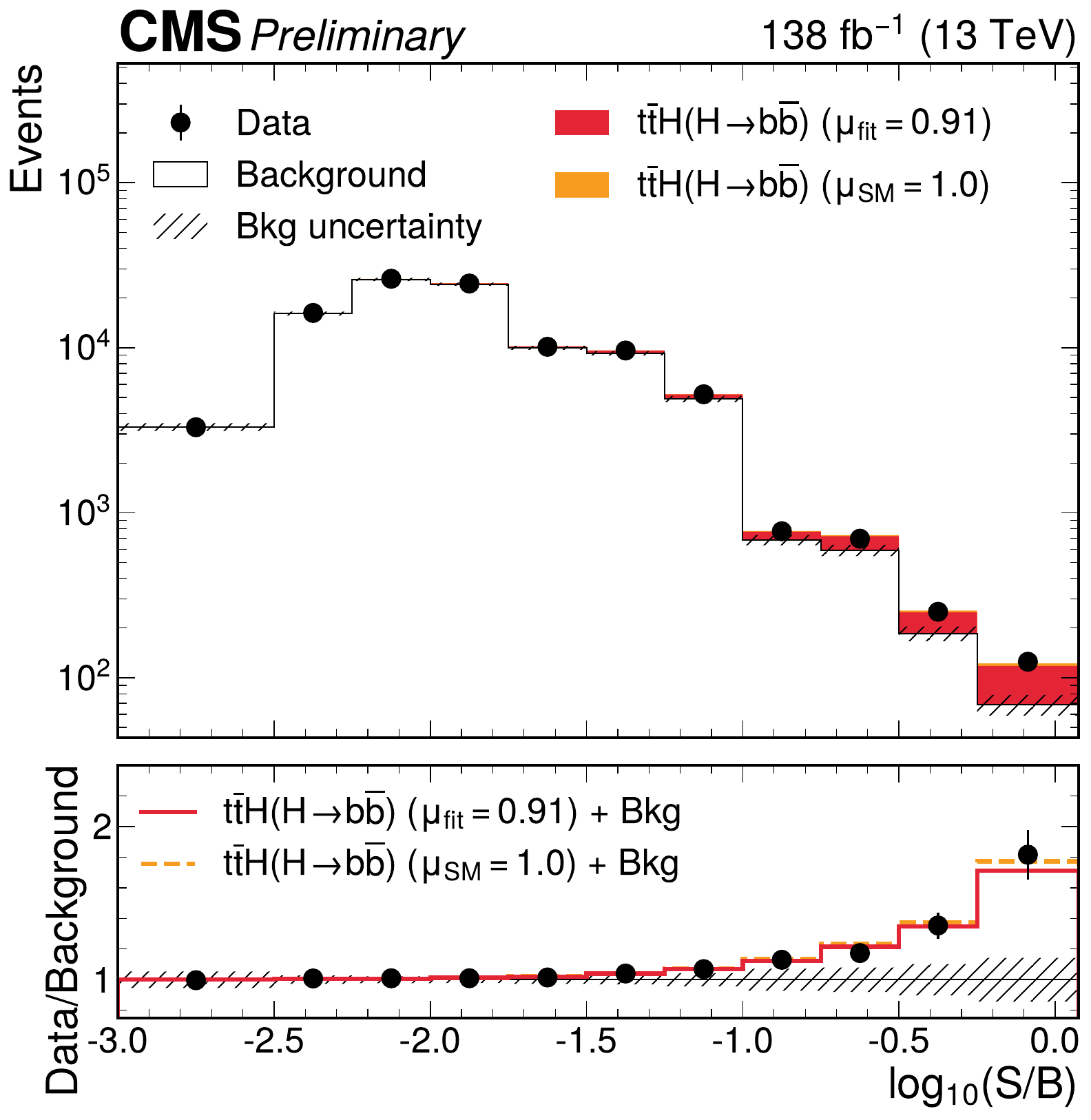}}
\end{minipage}
\caption[]{Observed and expected event yields from all SRs and CRs as a function of $\log_{10}(S/B)$, where $S$ are the expected \ttHcc (left) and \ttHbb (right) yields, and $B$ are the post-fit total background yields.
        Signal contributions are shown for the best-fit signal strength (red) and for the SM prediction, $\mu = 1$ (orange).
        The lower panel shows the ratio of the data to the post-fit background predictions~\cite{CMS-PAS-HIG-24-018}.}
\label{fig:soverb}
\end{figure}

Figure~\ref{fig:soverb} shows the observed and expected event yields from all CRs and SRs as a function of the logarithm of the ratio of \ttHcc (or \ttHbb) and background yields.
The best-fit signal strengths for \ttH production are determined to be $\muHcc = -1.6\pm4.5$ and $\muHbb = 0.91\pmasym{+0.26}{-0.22}$, with an observed (expected) significance of 4.4 (4.5) standard deviations for \ttHbb.
An upper limit on \muHcc is extracted using the CLs criterion, with an observed (expected) 95\% CL upper limit on \muHcc of 7.8 (8.7).

\begin{figure}
\begin{minipage}{0.45\linewidth}
\centerline{\includegraphics[width=0.99\linewidth]{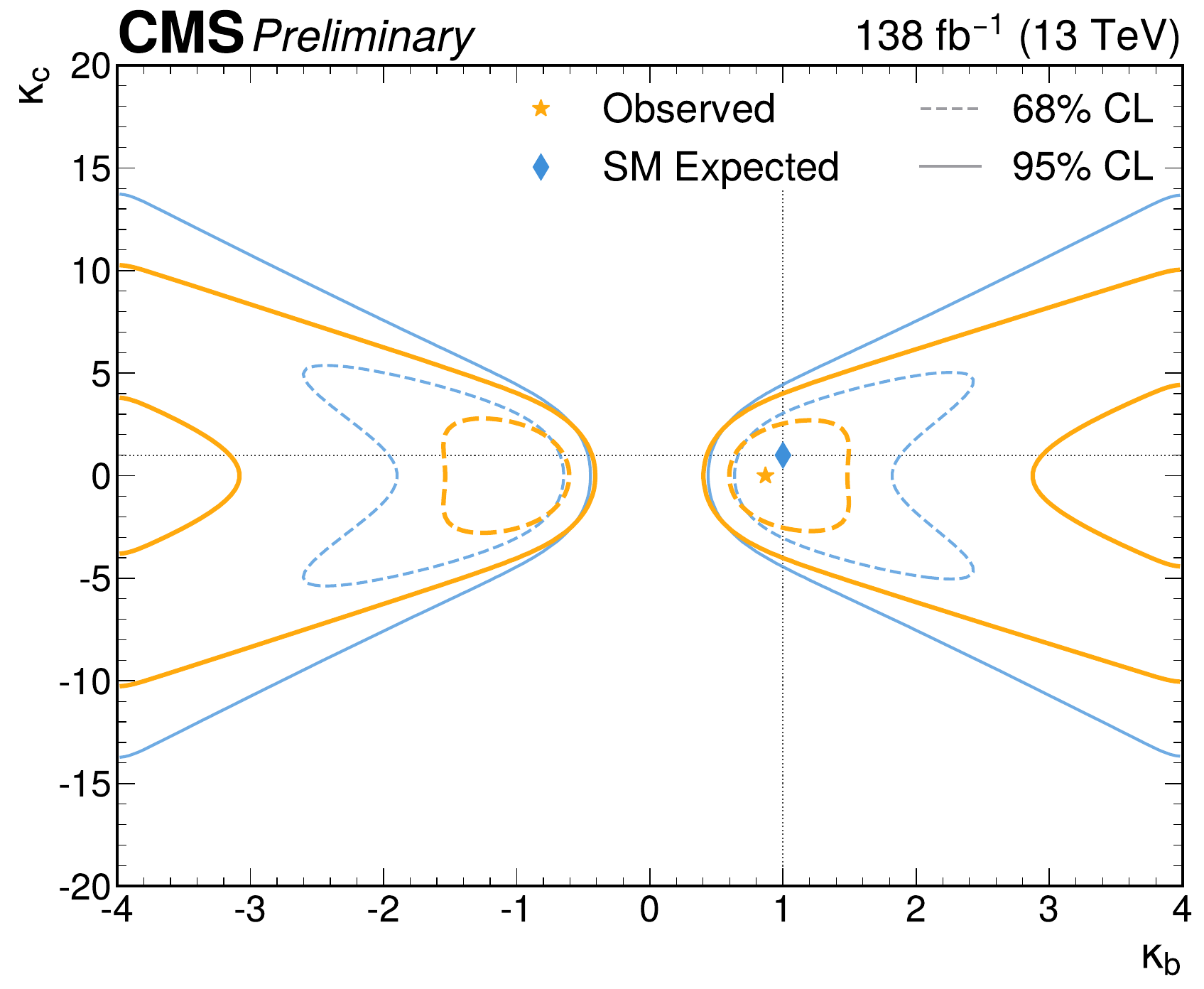}}
\end{minipage}
\hfill
\begin{minipage}{0.54\linewidth}
\centerline{\includegraphics[width=0.99\linewidth]{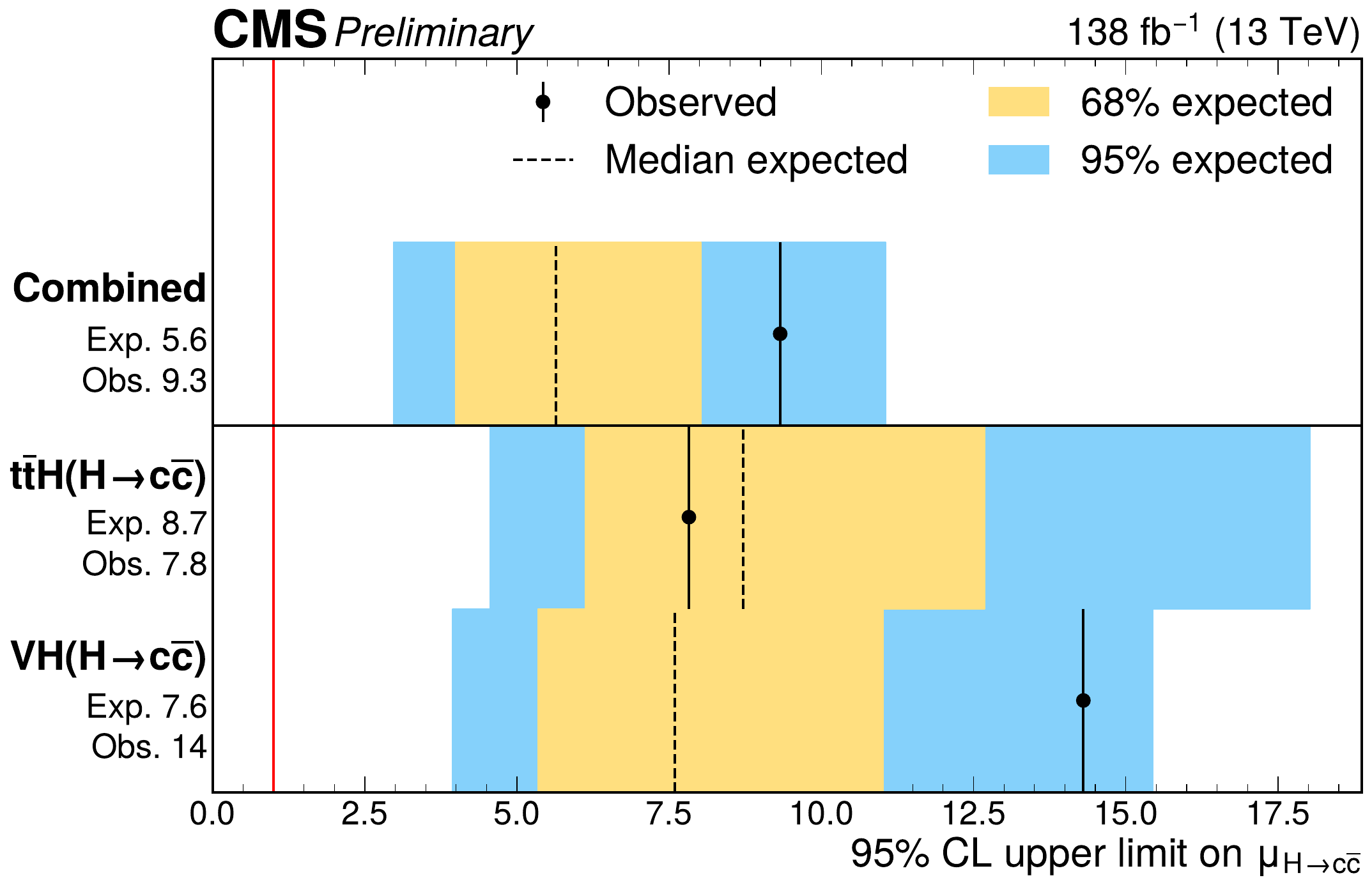}}
\end{minipage}
\caption[]{Constraints on the Higgs boson coupling modifiers \kappaC and \kappaB (left) and the 95\% CL upper limits on \muJustHcc (right)~\cite{CMS-PAS-HIG-24-018}.}
\label{fig:results}
\end{figure}

The result is interpreted in the $\kappa$-framework~\cite{deFlorian:2016spz} by parameterizing $\brHcc$ and $\brHbb$ in terms of the charm and bottom quark Yukawa coupling modifiers \kappaC and \kappaB.
Figure~\ref{fig:results} shows the two-dimensional profile likelihood scan of \kappaC and \kappaB.
When fixing \kappaB to the SM expectation of unity, the observed (expected) 95\% CL interval is $|\kappaC| < 3.0$ (3.3).

Finally, a combined analysis with the previous search in the \VH channel~\cite{CMS:HIG-21-008} is performed.
The observed (expected) 95\% CL upper limit on \muJustHcc, assuming SM production rates for \ttH and \VH, is 9.3 (5.6).
For \kappaC, the combination improves the expected $95\%$ CL interval to $|\kappaC|<2.7$, while the observed is $|\kappaC|<3.5$.
This represents the most stringent constraint on \kappaC to date.

\section*{References}
\bibliography{biblio}{}
\end{document}